\documentclass[twocolumn,showpacs,amsfonts,aps,prc,nofootinbib,floatfix,%
superscriptaddress]{revtex4}

\usepackage{amsmath}
\usepackage{bm}
\usepackage{graphicx}

\usepackage{epsfig}
\newcommand{\beq}{\begin{equation}}
\newcommand{\eeq}{\end{equation}}
\newcommand{\bea}{\vspace{0.25cm}\begin{eqnarray}}
\newcommand{\eea}{\end{eqnarray}}

\newcommand{\ro}{\mbox{{\boldmath
$\rho$}}}

\newcommand{\kb}{{{\bf k}}}

\def\lsim{\mathrel{\rlap{\lower4pt\hbox{\hskip1pt$\sim$}}
    \raise1pt\hbox{$<$}}}         
\def\gsim{\mathrel{\rlap{\lower4pt\hbox{\hskip1pt$\sim$}}
    \raise1pt\hbox{$>$}}}         

\newcommand{\landau}{L.D.~Landau Institute for Theoretical Physics,
        GSP-1, 117940, Kosygina Str. 2, 117334 Moscow, Russia}

\begin{document}


\title{
Multiplicity density at mid-rapidity 
in $AA$ collisions:
effect of meson cloud
}
\date{\today}

\author{B.G.~Zakharov}\affiliation{\landau}

\begin{abstract}
We study the influence of the meson cloud of the nucleon on
predictions of the Monte Carlo Glauber model for the charged particle 
multiplicity density at mid-rapidity in $AA$ collisions.
We find that for central $AA$ collisions
the meson cloud can increase the multiplicity density 
by $\sim 16-18$\%.
The meson-baryon Fock component 
reduces the required fraction of the binary collisions
by a factor of $\sim 2$ for Au+Au collisions at 
$\sqrt{s}=0.2$ TeV and  
$\sim 1.5$ for Pb+Pb collisions at 
$\sqrt{s}=2.76$ TeV.

\end{abstract}
%

\maketitle

\section{Introduction}
The understanding of the initial entropy/energy distribution  is crucial for
hydrodynamical simulation of the evolution of the hot quark-gluon plasma 
(QGP) in high-energy $AA$ collisions. A rigorous determination of the initial
conditions  for the plasma fireball in $AA$ collisions is presently impossible.
The most popular methods in use for this purpose at the present time
are the IP-Glasma model \cite{IP-GL,IP-GL2} and the wounded nucleon Glauber 
model \cite{WNG,KN}. The IP-Glasma approach is based on 
the pQCD color-glass 
condensate model \cite{CGC}. Unfortunately the applicability of the pQCD 
in the IP-Glasma model is questionable since it assumes that gluon fields
can be treated perturbatively down to an infrared scale $m\sim \Lambda_{QCD}$  
\cite{IP-GL,IP-GL2}.
It is several times smaller than 
the inverse gluon correlation radius in the QCD vacuum $1/R_c\sim 0.75$ GeV
\cite{Shuryak}, which is the natural lower limit for the virtuality scale of the perturbative gluons.
In the dipole approach \cite{NZ12} for
$m \sim 0.75$ GeV the perturbative contribution to the hadronic cross sections 
turns out to be smaller than the nonperturbative one up to 
$\sqrt{s}\sim 10^3$ GeV \cite{Zoller2}. For this reason even at the 
LHC energies a purely perturbative  treatment of the hadron 
cross sections is questionable.

The wounded nucleon Glauber model \cite{WNG,KN} is a 
phenomenological scheme.
Originally \cite{WNG} it was assumed that in $AA$ 
collisions each nucleon undergoing inelastic soft interaction
(participant) produces a fixed contribution to the multiplicity
rapidity density. 
At mid-rapidity  ($\eta=0$) in the c.m. frame  this contribution
equals half of the $pp$ multiplicity
rapidity density. It gives for $AA$ collisions 
the multiplicity density $\propto N_{part}$, where $N_{part}$ is the number
of participants in both the colliding nuclei. 
Later, in \cite{KN} it was proposed to include in the model
the contribution  
from hard processes that gives 
the particle density proportional to the number of the binary 
collisions $N_{coll}$. In this two component version the charged particle 
multiplicity density in 
$AA$ collisions takes the form
\beq
\frac{dN_{ch}(AA)}{d\eta}=\frac{(1-\alpha)}{2}n_{pp}N_{part}+
\alpha n_{pp}N_{coll}\,,
\label{eq:10}
\eeq
where $n_{pp}=dN_{ch}/d\eta$ is the multiplicity density in $pp$ collisions,
and $\alpha$ characterizes the magnitude of hard processes
to multiparticle production. 
In the Glauber model 
model $N_{part}$ and $N_{coll}$ can be expressed via the inelastic $pp$ cross 
section and the nuclear density.
Fitting the data on the 
centrality dependence of the charged particle multiplicity
in Au+Au collisions at $\sqrt{s}=0.2$ TeV and in Pb+Pb collisions
at 2.76 TeV gives $\alpha\approx 0.13-0.15$
\cite{PHOBOS0405,STAR0808,LHC_MCGL}. For such a value of $\alpha$ the hard
contribution to the particle production in $AA$ collisions turns out to 
be rather  
large ($\sim 40-50$\% for central collisions). 
It is important that the two component Glauber model allows the 
Monte Carlo formulation \cite{PHOBOS_MC,GLISS,GLISS2}. The Monte Carlo Glauber 
(MCG) model has proved to be a useful tool for analysis of the event-by-event 
fluctuations of observables in $AA$ collisions.

The model of wounded nucleons has been also formulated at quark level
\cite{woun_q1,woun_q2} when inelastic interaction of the nucleon
is treated as a combination of inelastic interactions of its constituent
quarks. However, in this picture the quark contribution to the multiplicity
required for description of data on $AA$ collisions may differ substantially
from the one that is necessary for $pp$ collisions \cite{Bozek_2016_3q}.
Say,  the data on Au+Au collisions at $\sqrt{s}=0.2$ TeV 
require the quark contribution suppressed by a factor $\sim 1.4$ as 
compared to $pp$ interaction \cite{Bozek_2016_3q}. However, the 
situation with consistency between $AA$ and $pp$ collisions
becomes better if the nucleon is treated as a quark-diquark system
\cite{Bozek_qD}. 

The common feature of the wounded nucleon models with internal nucleon
structure is the nonlinear increase of $dN_{ch}(AA)/d\eta$ 
with
the number of wounded nucleons even without the hard contribution
\cite{Voloshin_3q,Bozek_qD,Bozek_2016_3q,Loizides_q}.
This is due to the growth of the fraction of the wounded constituents in
each nucleon in $AA$ collisions as compared to that in $pp$ collisions. 
It is clear that a similar effect should arise 
from the meson cloud of the nucleon.
The total weight of the meson-baryon Fock states 
in the nucleon may be as large as $\sim 40$\% \cite{ST}
(with the dominant contribution from the $\pi N$ component).
The purpose of the present work is to study within the MCG approach 
the possible effect of the meson-baryon component of the nucleon on the
multiplicity rapidity density in $AA$ collisions.
We will analyze within the MCG model with the meson cloud  
data on Au+Au collisions at $\sqrt{s}=0.2$ \cite{STAR1} and Pb+Pb collisions
at $2.76$ TeV \cite{ALICE1}. 
 
\section{Theoretical framework}
At high energies the wave function of the physical nucleon
becomes identical to that 
in the infinite momentum 
frame (IMF). It can be written in the form \cite{ST,Zoller_MB}
\beq
|N\rangle_{phys}=\sqrt{1-n_{MB}}|N\rangle+
\sum_{MB}\int dxd\kb\Psi_{MB}(x,\kb)|MB\rangle\,,
\label{eq:20}
\eeq
where $N$, $B$, and $M$ denote the bare baryon and meson states, 
$x$ is the fractional longitudinal meson momentum in the physical nucleon, $\Psi_{MB}$ is
the probability amplitude for the $MB$ Fock state, and
\beq
n_{MB} =\sum_{MB}\int dxd\kb|\Psi_{MB}(x,\kb)|^2
\label{eq:30}
\eeq
is the total weight of the $MB$ Fock components.
The energy denominator of time-ordered perturbation theory in the IMF 
for the $MB$ component  reads
$E_N-E_M-E_B\approx [m_N^2-M^{2}_{MB}(x,\kb^2)]/2E_N$, 
where
\beq
M^{2}_{MB}(x,\kb^2)=\frac{m_M^2+\kb^2}{x}+\frac{m^2_{B}+\kb^2}{1-x}
\label{eq:40}
\eeq
is the squared invariant mass of the $MB$ system. 
For the normalization  corresponding to (\ref{eq:30})
the IMF wave function (for point-like particles) may be written as 
\beq
\Psi_{MB}(x,\kb)=
\frac{\langle MB|V|N\rangle}
{4\pi^{3/2}\sqrt{x(1-x)}\left[m_N^2-M^2_{MB}(x,\kb)\right]}\,.
\label{eq:50}
\eeq
Here $\langle MB|V|N\rangle$ is the vertex factor in the IMF-limit, which
depends on the form of the Lagrangian. For the dominant 
$\pi N$ state $\langle \pi N'|V|N\rangle=g_{\pi NN}\bar{u}_{N'}\gamma_5u_{N}$  
(the
helicity dependent vertex functions for different $MB$ states
can be found in \cite{ST}). In phenomenological applications
the internal structure of the hadrons 
is accounted for by multiplying the vertex 
factor for point-like particles by a form factor, $F$, which
in the IMF scheme depends on $x$ and $\kb$ only via 
$M_{MB}(x,\kb)$ \cite{Zoller_MB,ST,MST_MB}.

To a good approximation on can account for in (\ref{eq:20}) only  
$\pi N$, $\pi\Delta$, $\rho N$ and $\rho \Delta$ two-body states 
\cite{ST,MST_MB}. 
Since the bare $\Delta$ and $\rho$-meson states have the same quark content
as $N$ and $\pi$, it is reasonable to assume that 
their inelastic interactions 
are similar to that for $N$ and $\pi$ states. 
Then, from the point of view of the MCG model, each physical nucleon 
interacts with the probability $1-n_{MB}$ as the bare $N$ and with the 
probability $n_{MB}$ as the two-body $\pi N$ system.
For $\pi$-meson the $x$-distribution is peaked at $x\sim 0.3$ and for
$\rho$-meson at $x\sim 0.5$ \cite{ST}. For simplicity in our MCG calculations 
we neglect fluctuations of $x$ and take $x=0.3$.
For the transverse spacial distribution of the $MB$ state
we use the distribution of the dominant $\pi N$ component.
It was renormalized to match the total weight of the $MB$ component 
$n_{MB}=0.4$ \cite{ST}.  
We calculated the transverse spacial distribution using the 
dipole formfactor \cite{MST_MB} 
\beq
F=\left(\frac{\Lambda^2+m_N^2}{\Lambda^2+M^{2}_{\pi N}(x,\kb)}\right)^2\,.
\label{eq:60}
\eeq
We take $\Lambda=1.3$ GeV, such a value is supported by the data on 
$pp\to nX$ \cite{HSS_MB}. 
It gives for the mean squared transverse radius of the $MB$ component 
$\langle \rho^2_{MB}\rangle^{1/2}\approx0.87$ fm. 
However, the results 
of the MCG simulation depends weakly on the value of $\Lambda$.
This is due to the fact that in the Glauber
model there is no shadowing effect for inelastic interactions.

In our model inelastic interaction of the physical nucleons
from the colliding objects is a combination of $N+N$, $N+MB$, $MB+N$ and
$MB+MB$ interactions. We assume that the inelastic cross sections
for the bare states obey the constituent quark counting rule
$4\sigma^{NN}_{in}=6\sigma^{MB}_{in}=9\sigma^{MM}_{in}$.
For the profiles of the probability of $ab$ inelastic interaction 
in the impact parameter we use a Gaussian
form 
\beq
P_{ab}(\rho)=\exp\left(-\pi \rho^2/\sigma_{in}^{ab}\right)\,.
\label{eq:70}
\eeq
The value of $\sigma_{in}^{NN}$ has been adjusted to reproduce 
the experimental inelastic $pp$ cross section $\sigma_{in}^{pp}$ (see below).

We consider the multiplicity density at $\eta=0\,\,(y=0)$.
The direct data on $dN_{ch}/d\eta$ for pion-proton and pion-pion 
collisions for RHIC-LHC energies are absent. 
Calculations within the quark-gluon string
scheme \cite{Kaidalov,Capella} show that the charged particle 
multiplicity density in the central rapidity region for pion-proton and
pion-pion collisions is somewhat bigger than for proton-proton collisions. 
To good accuracy this excess compensates a possible reduction of 
the multiplicity
density in $\pi N$ and $\pi \pi$ interactions due to somewhat smaller 
c.m. energy in our model.
For this reason we assume that all the wounded bare particles produce the same
amount of entropy per unit pseudorapidity $\eta$ in the 
c.m. frame of colliding objects ($pp$ or $AA$).
We ignore the effect of a small rapidity shift ($\sim 0.5$) of the c.m. frame 
for pairs with different energies (as occurs for $\pi N$ interactions)
on the entropy rapidity density since it
is flat at mid-rapidity.

The total rapidity density for $AA$ collisions
is 
the sum of the contributions from the sources corresponding to 
the wounded constituents and to the binary collisions 
\beq
\frac{dS}{dy}=\sum_{i=1}^{N_w} \frac{dS_w^{i}}{dy}+
\sum_{i=1}^{N_{bin}} \frac{dS_{bin}^{i}}{dy}\,.
\label{eq:80}
\eeq
We write the contribution of each source from the 
wounded constituents
as $dS_w^i/dy=\frac{(1-\alpha)}{2}S$.
The contribution of each binary collision is $dS_{bin}^i/dy=S$, and 
for each pair of wounded particles the 
probability of a hard binary collision is $\alpha$. 
We assume an isentropic expansion of the QGP. In this case 
the initial entropy rapidity density is proportional the charged
particle pseudorapidity density $dS/dy=C dN_{ch}/d\eta$,
where $C\approx 7.67$ \cite{BM-entropy}. In this approximation
one can replace in (\ref{eq:80}) the entropy density 
by the pseudorapidity charged particle density. 
And the fluctuating entropy density $S$ for each source 
is replaced by the fluctuating 
pseudorapidity charged particle density $n=S/C$.
We describe the fluctuations of $n$ by the Gamma distribution
\beq
P(n,\langle n\rangle)=
\left(\frac{n}{\langle n\rangle}\right)^{k-1}
\frac{k^k\exp\left[-nk/\langle n\rangle\right]}
{\langle n\rangle \Gamma(k)}\,.
\label{eq:90}
\eeq
The parameters $\langle n\rangle$ and $k$ have been fitted from
$pp$ data on $dN_{ch}/d\eta$ (see below).
Note however that for $AA$ 
collisions the results are only weakly sensitive to 
fluctuations of $n$ (except for the region of very high multiplicities).

For calculation of the centrality dependence of the charged particle
multiplicity in $AA$ collisions the distribution of the entropy
rapidity density in the transverse coordinates, $\rho_s=dS/dyd\ro$,
is not important. However, it is necessary for calculation of geometric
quantities such as the initial anisotropy $\epsilon_n$ 
\cite{Teaney_en}
\beq
\epsilon_n=\frac{\int d\ro \rho^n e^{in\phi}\rho_s(\ro)}
{\int d\ro \rho^n\rho_s(\ro)}\,.
\label{eq:100}
\eeq
In the approximation of the point-like sources 
we have
\beq
\rho_s(\ro)
=\sum_{i=1}^{N_w} \delta(\ro-\ro_i)\frac{dS_w^{i}}{dy}+
\sum_{i=1}^{N_{bin}} \delta(\ro-\ro_i)\frac{dS_{bin}^{i}}{dy}\,.
\label{eq:110}
\eeq
We assume that for each binary collision the source
is located in the middle between 
colliding constituents. 
Physically the approximation of the point-like sources is
clearly unreasonable. To account for qualitatively the finite size
of the sources we replaced in our MCG code the $\delta$ functions
in (\ref{eq:110}) by a Gaussian distribution  
$\exp{\left(-\ro^2/a_s^2\right)}/\pi a_s^2$ with $a_s=0.7$ fm.
We observed that the results for the anisotropy coefficients 
$\epsilon_n$ becomes sensitive to the smearing of the sources 
only for very peripheral collisions.

We perform calculations using the Woods-Saxon nuclear distribution
\beq
\rho_{A}(r)=\frac{c}{1+\exp[(r-R_A)/a]}\,,
\label{eq:120}
\eeq
where $c$ is the normalization constant, 
$R_{A}=(1.12A^{1/3}-0.86/A^{1/3})$ fm, $a=0.54$ fm \cite{GLISS2}.

\section{Numerical results}
In numerical calculations
we take $n_{pp}=2.65$ at $\sqrt{s}=0.2$ TeV
obtained by the UA1 collaboration \cite{UA1_pp}. 
The direct $pp$ data on $n_{pp}$ at 
$\sqrt{s}=2.76$ TeV are absent. 
We obtained it with the help of the power law interpolation
between the CMS data at $\sqrt{s}=2.36$ TeV \cite{CMS0621}
($n_{pp}=4.47\pm 0.04(\mbox{stat.})
\pm 0.16 (\mbox{syst.})$)  and 
at $\sqrt{s}=7$ TeV \cite{CMS3299}   
($n_{pp}=5.78\pm 0.01(\mbox{stat.}) \pm 0.23 (\mbox{syst.})$). 
It gives $n_{pp}\approx
4.65$ at $\sqrt{s}=2.76$ TeV.
The multiplicity densities measured in 
\cite{UA1_pp,CMS0621,CMS3299} correspond to the non-single-diffractive
(NSD) events. For this reason in the MCG simulation 
one should also take for $\sigma_{in}^{pp}$ the cross section  
corresponding to the NSD event class.
The exclusion of the diffractive contribution to the inelastic
cross section is reasonable since the diffractive events
do not contribute to the mid-rapidity multiplicity density.

We use for the NSD $pp$ cross section
at $\sqrt{s}=0.2$ TeV
the value $35$ mb measured by the UA1 collaboration \cite{UA1_pp},
and  at $\sqrt{s}=2.76$ TeV
the value $50.24$ mb obtained by the ALICE collaboration
\cite{ALICE4968}.
Making use of the above values of the NSD $\sigma_{in}^{pp}$
we fitted $\sigma_{in}^{NN}$. 
For the scenario with meson cloud we obtained 
\beq
\sigma_{in}^{NN}[\sqrt{s}=0.2\,,2.76\,\mbox{TeV}] \approx [26.15,\,38.4]
\,\,\mbox{mb}\,.
\label{eq:130}
\eeq
In the scenario without meson cloud $\sigma_{in}^{NN}$ is equal simply 
to  the experimental NSD $pp$ cross section.
The parameters $\langle n\rangle$ and $k$ in the Gamma distribution 
(\ref{eq:90}) have 
been fitted to reproduce the experimental $n_{pp}$ and to satisfy the
relation $n_{pp}/D=1$ ($D^2$ is a variance of $dN_{ch}/d\eta$) which 
is well satisfied for the experimental multiplicity distribution
in the pseudorapidity window $|\eta|<0.5$ at $\sqrt{s}=0.2$ TeV 
\cite{UA1_pp} and at $\sqrt{s}=2.36$ TeV \cite{ALICE_pp2.36}.
For the scenario without meson cloud $\langle n\rangle$
equals the experimental $n_{pp}$ for any fraction of the binary 
collisions. For $\alpha=0$ the relation $n_{pp}/D=1$ gives $k=0.5$.
For nonzero $\alpha$ the value of $k$ grows weakly with $\alpha$, 
but the deviation
from $0.5$ is small. For the scenario with meson cloud the required
value of $\langle n\rangle$ is smaller than $n_{pp}$, and $k$
is close to $0.5$.  

We first fitted the parameters $\langle n\rangle$ and $k$ to 
the $pp$ data for set of $\alpha$. Then we used them to fit the parameter 
$\alpha$ to best reproduce the
data on the centrality dependence of $dN_{ch}/d\eta$
in Au+Au collisions at $\sqrt{s}=0.2$ TeV from STAR \cite{STAR1}
and in Pb+Pb collisions at $\sqrt{s}=2.76$ TeV from ALICE \cite{ALICE1}.
For Au+Au collisions at $\sqrt{s}=0.2$ TeV we obtained 
$\alpha\approx0.06$ and $\alpha\approx 0.135$  for the scenarios 
with and without meson cloud, respectively.
And for Pb+Pb collisions at $\sqrt{s}=2.76$ TeV 
for these two scenarios our fits give
$\alpha\approx0.09$ and $\alpha\approx 0.14$.
For the above values of $\alpha$ the parameters of the Gamma distribution
(\ref{eq:90}) obtained from the fit with meson 
cloud to the $pp$ data  read
\beq
\langle n\rangle[\sqrt{s}=0.2,\,2.76\,\text{TeV}]\approx
[2.39,\,4.13]\,,
\label{eq:140}
\eeq
\beq
k[\sqrt{s}=0.2,\,2.76\,\text{TeV}]\approx
[0.506,\,0.52]\,.
\label{eq:150}
\eeq
For the scenario without meson cloud $\langle n\rangle=n_{pp}$, and
\beq
k[\sqrt{s}=0.2,\,2.76\,\text{TeV}]\approx
[0.57,\,0.57]\,.
\label{eq:160}
\eeq
As expected, accounting for the meson cloud leads to a reduction of the
required fraction of the binary collisions. The effect becomes 
smaller at the LHC energy. It is due to an increase of the 
interaction radius from RHIC to LHC, resulting in the lower sensitivity 
to the internal nucleon structure at the LHC energy.

In Figs.~1,~2 we compare our calculations for the fitted values of $\alpha$ 
with STAR \cite{STAR1} and ALICE \cite{ALICE1} data. The theoretical 
histograms have been obtained by Monte Carlo generation of  $\sim 2\times 10^6$
events.
To illustrate the magnitude of the effect of the meson cloud
in Figs.~1a and 2a we show the results for the scenario without meson 
cloud obtained with $\alpha$ for the scenario with meson cloud.
One can see that at small centrality the meson cloud increases
the multiplicity by $\sim 16-18$\%.
Note that our calculations do not assume a certain
internal structure of the bare baryon and meson states.
For this reason one can expect that the 
long range meson-baryon fluctuations in the nucleon wave function
should increase the multiplicity in $AA$ collisions 
in any scheme. 

We also studied the effect of meson cloud on the 
initial anisotropy coefficients $\epsilon_n$ ($n=2-5$).
We found that the effect of the meson cloud is small
(except for very peripheral collisions where the results are
not robust due to their sensitivity to the entropy distribution
for the wounded constituents and the binary collisions).
Recently there was interest in the multiplicity dependence of the 
ellipticity $\epsilon_2$ for U+U collisions 
\cite{UU_Voloshin,UU_Bass,UU_Singh}.
In \cite{UU_Voloshin}
it was predicted that due to prolate shape of the $^{238}$U nucleus
the initial $\epsilon_2$ should has a knee structure at multiplicities
in the top 1\% U+U collisions related to the growth of the contribution of 
the binary collisions for the tip-tip configurations of the colliding nuclei.  
But the elliptic flow $v_2$ measured by STAR \cite{UU_Pandit,UU_STAR} in
U+U collisions at $\sqrt{s}=193$ GeV
shows no indication of a knee structure. This challenged the picture with a 
significant
contribution of the binary collisions, and stimulated study of 
alternative ansatze for the entropy deposition in the Glauber picture
\cite{UU_Bass,UU_Singh}.
However, the Glauber calculations of \cite{UU_Voloshin} have 
been performed neglecting the fluctuations of the multiplicity in $NN$
collisions. Later in \cite{UU_Broniowski} it was demonstrated that
the knee structure vanishes when the fluctuations are taken into account.
Our calculations for U+U collisions also show that the knee structure
in $\epsilon_2$ 
is swept out (both with and without
meson cloud) when the fluctuations of the sources are taken
into account.

\begin{figure}[ht]
\epsfig{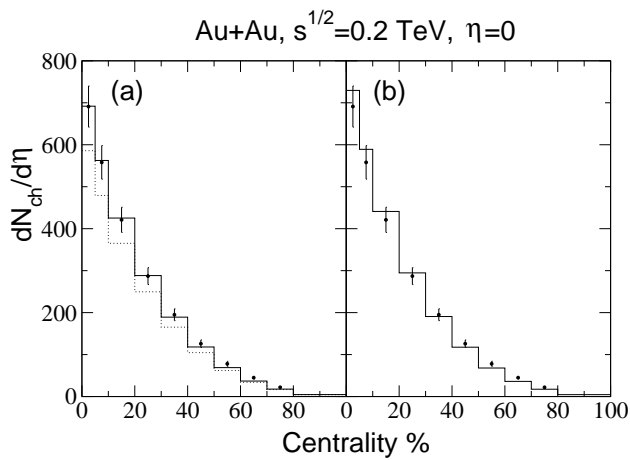}
\caption{\small Centrality dependence of $dN_{ch}/d\eta$ for 
Au+Au collisions at $\sqrt{s}=0.2$ TeV. Left: 
MCG simulation for the scenario with (solid) and without (dotted) meson cloud
for $\alpha=0.06$.
Right: MCG simulation for the scenario without meson cloud
for $\alpha=0.135$. Data are from STAR \cite{STAR1}.
}
\end{figure}

\begin{figure}
\epsfig{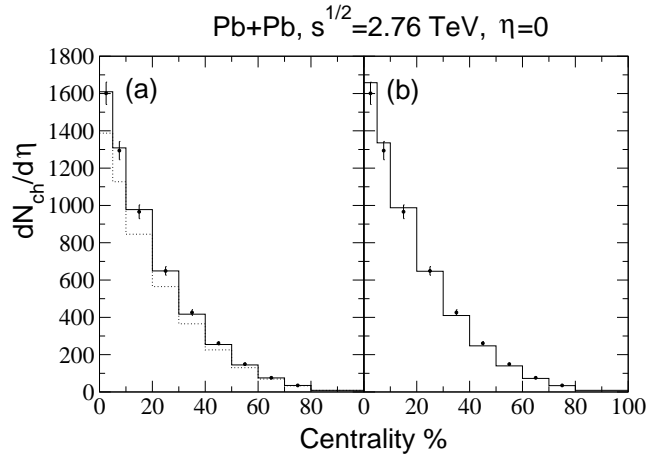}
\caption{\small Centrality dependence of $dN_{ch}/d\eta$ for 
Pb+Pb collisions at $\sqrt{s}=2.76$ TeV. Left: 
MCG simulation for the scenario with (solid) and without (dotted) meson cloud
for $\alpha=0.09$.
Right: MCG simulation for the scenario without meson cloud
for $\alpha=0.14$. Data are from ALICE \cite{ALICE1}.
}
\end{figure}

\section{Summary}
We have studied the influence of the meson cloud on
predictions of the MCG model for $AA$ collisions.
We find that for central $AA$ collisions
the meson cloud can increase the multiplicity density in
the central rapidity region by $\sim 16-18$\%.
Accounting for the meson-baryon Fock components
of the nucleon reduces the required
fraction of the binary collisions
by a factor of $\sim 2$ for Au+Au collisions at 
$\sqrt{s}=0.2$ TeV and  
$\sim 1.5$ for Pb+Pb collisions at 
$\sqrt{s}=2.76$ TeV.
One can expect that the observed increase of the multiplicity 
in $AA$ collisions due to the virtual meson-baryon states
in the physical nucleon should exist 
in other models for the initial conditions in $AA$ collisions.

\begin{acknowledgments} 	
I thank 
W.~Broniowski and 
S.A.~Voloshin for communications.
This work is supported 
in part by the 
grant RFBR
15-02-00668-a.
\end{acknowledgments}

\end{document}